\documentclass[preprint,superscriptaddress,nofootinbib]{revtex4}
  
  \usepackage{amssymb,amsmath,graphicx,subfigure,color}
  \usepackage[colorlinks]{hyperref}
  \allowdisplaybreaks[4]
  \begin{document}

  \title{Possibility of experimental study on nonleptonic
         $B_{c}^{\ast}$ weak decays}
  \author{Yueling Yang} 
  \affiliation{Institute of Particle and Nuclear Physics,
              Henan Normal University, Xinxiang 453007, China}
  \author{Liting Wang} 
  \affiliation{Institute of Particle and Nuclear Physics,
              Henan Normal University, Xinxiang 453007, China}
  \author{Jinshu Huang} 
  \affiliation{School of Physics and Electronic Engineering,
              Nanyang Normal University, Nanyang 473061, China}
  \author{Qin Chang} 
  \affiliation{Institute of Particle and Nuclear Physics,
              Henan Normal University, Xinxiang 453007, China}
  \author{Junfeng Sun} 
  \affiliation{Institute of Particle and Nuclear Physics,
              Henan Normal University, Xinxiang 453007, China}

   \begin{abstract}
   The ground vector $B_{c}^{\ast}$ meson has not yet been
   experimentally discovered until now.
   Besides the dominant electromagnetic decays,
   nonleptonic weak decays provide another
   choice to search for the mysterious $B_{c}^{\ast}$ mesons.
   Inspired by the potential prospects of $B_{c}^{\ast}$
   mesons in future high-luminosity colliders,
   nonleptonic $B_{c}^{\ast}$ weak decays induced by bottom
   and charm quark decays are studied within the SM by using
   a naive factorization approach.
   It is found that for $B_{c}^{\ast}$ ${\to}$ $B_{s,d}{\pi}$,
   $B_{s,d}^{\ast}{\pi}$, $B_{s,d}{\rho}$, $B_{s}K$,
   $B_{s}^{\ast}K$, $B_{s}K^{\ast}$, ${\eta}_{c}(1S,2S){\pi}$,
   ${\eta}_{c}(1S,2S){\rho}$ and ${\psi}(1S,2S){\pi}$ decays,
   a few hundred and even thousand of events might be
   observable in CEPC, FCC-ee and LHCb@HL-LHC experiments.

   \href{https://doi.org/10.1088/1674-1137/aca00d}
        {Chin. Phys. C 47, 013110 (2023).}
   \end{abstract}

  \maketitle

   \section{Introduction}
   \label{sec01}
   According to the $q\bar{q}$ quark model assignments for
   mesons, the bottom-charmed mesons are unique particles
   consisting of two heavy quarks with different flavors,
   because it is generally assumed that
   the top quark being the
   heaviest element fermion of the standard model (SM)
   has a very short lifetime and decays before hadronization.
   The bottom-charmed mesons are isospin singlets, and have
   nonzero additive quantum numbers, $Q$ $=$ $B$ $=$ $C$
   $=$ ${\pm}1$, where $Q$, $B$ and $C$ are respectively
   electric charge, bottom and charm.
   The bottom-charmed mesons, analogical with hidden-flavored
   charmonium $c\bar{c}$ and bottomonium $b\bar{b}$,
   are often considered as  nonrelativistic bound states.
   The hyperfine splitting interactions divide the
   ground bottom-charmed states into the spin singlet
   $1^{1}S_{0}$ and triplet $1^{3}S_{1}$, corresponding to
   the pseudoscalar $B_{c}$ meson with the spin-parity $J^{P}$ $=$
   $0^{-}$ and vector $B_{c}^{\ast}$ meson with $J^{P}$ $=$
   $1^{-}$, respectively.

   The pseudoscalar $B_{c}$ meson was first discovered via
   $B_{c}$ ${\to}$ $J/{\psi}{\pi}$ and
   $J/{\psi}{\pi}{\pi}^{+}{\pi}^{-}$ decay modes with
   hadronic $Z^{0}$ decay sample data collected
   by the DELPHI detector at the $e^{+}e^{-}$ collider LEP
   in 1997 \cite{Phys.Lett.B.398.207}.
   The natural properties of the
   $B_{c}$ meson including its mass,
   lifetime, and spin-parity have now been
   well determined \cite{pdg2022}.
   In the meantime, the vector $B_{c}^{\ast}$ meson has not been
   definitively identified by experimental physicists yet.
   All the information on the
   $B_{c}^{\ast}$ meson come only from
   theoretical calculation for the moment.
   Undoubtedly, it is a matter of great urgency to find and
   identify the $B_{c}^{\ast}$ meson in experiments, which is the
   basis and prerequisite to obtine a deeper insight into
   the desired $B_{c}^{\ast}$ meson, and distinguish different
   theoretical models.

   The production probability of bottom-charmed mesons is
   far less than that of heavy-light $B_{u,d}$ mesons,
   and heavy-heavy charmonium and bottomonium.
   First, two heavy quark pairs of both $b\bar{b}$ and
   $c\bar{c}$ should be almost simultaneously produced
   in a high energy process, and then the $b$ (or $\bar{b}$)
   quark from a $b\bar{b}$ pair and $\bar{c}$ (or $c$) quark
   from a $c\bar{c}$ pair should be lucky enough to
   combine and finally form bottom-charmed
   mesons \cite{ijmpa.12.4039}.
   In positron-electron collisions, the bottom-charmed
   mesons can be produced via the $Z^{0}$ boson decay,
     \begin{equation}
     e^{+} + e^{-}\, {\to}\, Z^{0}\, {\to}\,
     (b\bar{c})+\bar{b}+c
     \label{epem-z0-bc},
     \end{equation}
   for example, the observation of $B_{c}$ mesons by
   DELPHI \cite{Phys.Lett.B.398.207},
   ALEPH \cite{Phys.Lett.B.402.213} and
   OPAL \cite{Phys.Lett.B.420.157} detectors at
   LEP experiments.
   In hadron-hadron collisions, the bottom-charmed
   mesons can be produced via gluon-gluon fusion and
   quark-antiquark annihilation,
     \begin{equation}
     g + g {\to}\, Z^{0}\, {\to}\,
     (b\bar{c})+\bar{b}+c
     \label{gg-z0-bc},
     \end{equation}
     \begin{equation}
     q + \bar{q}\, {\to}\, Z^{0}\, {\to}\,
     (b\bar{c})+\bar{b}+c
     \label{qq-z0-bc},
     \end{equation}
   for example, the observation of $B_{c}$ meson by
   CDF \cite{PhysRevLett.100.182002} and
   D0 \cite{PhysRevLett.101.012001} at Fermilab Tevatron,
   by LHCb \cite{JHEP.2020.07.123} at LHC experiments.
   More and more $B_{c}^{\ast}$ mesons are expected to be
   accessible in the future.
   Given the branching ratio of $Z^{0}$ boson decay
   ${\cal B}r(Z^{0}{\to}b\bar{b})$
   $=$ $15.12(5)\%$ \cite{pdg2022} and the bottom quark
   fragmentation fraction $f(b{\to}B_{c}^{\ast})$ ${\sim}$
   $6{\times}10^{-4}$ \cite{npa.953.21,cpc.43.083101,
   PhysRevD.100.034004},
   there will be more than $10^{8}$ $B_{c}^{\ast}$ mesons from
   $10^{12}$ $Z^{0}$ boson decays at the Circular Electron
   Positron Collider (CEPC) \cite{cepc},
   and more than $10^{9}$ $B_{c}^{\ast}$ mesons from $10^{13}$
   $Z^{0}$ boson decays at the Future Circular
   Collider (FCC-ee) \cite{fcc}.
   Assuming the $B_{c}^{\ast}$ production cross section is
   about $100$ nb for $pp$ collisions at $\sqrt{s}$ $=$ $13$
   TeV \cite{PhysRevD.97.114022}, more than $3{\times}10^{10}$
   $B_{c}^{\ast}$ mesons will be available with a total
   integrated luminosity of 300 ${\rm fb}^{-1}$ at LHCb
   of HL-LHC \cite{1808.08865}.
   The huge amount of experimental data provides an excellent
   opportunity and solid foundation to carefully study
   the $B_{c}^{\ast}$ mesons.

   Because of the nonrelativistic nature, the mass of the
   $B_{c}^{\ast}$ meson is approximately equal to the
   sum of the mass of bottom and charm quarks, {\em i.e.},
   $m_{B_{c}^{\ast}}$ ${\simeq}$ $m_{b}$ $+$ $m_{c}$.
   A recent lattice calculation gives $m_{B_{c}^{\ast}}$
   $=$ $6331(7)$ MeV \cite{PhysRevLett.121.202002},
   which agrees basically with many other theoretical
   estimations with various theoretical models ({\em e.g.},
   references [150-207] of Ref. \cite{Eur.Phys.J.C.81.1110}).
   Owing to the hierarchical relationships among mesonic
   mass, $m_{B_{c}^{\ast}}$ $<$ $m_{B}$ $+$ $m_{D}$ and
   $m_{B_{c}^{\ast}}$ $<$ $m_{B_{c}}$ $+$ $m_{\pi}$,
   the $B_{c}^{\ast}$ meson can not decay through the
   strong interactions.
   The dominant decay of the $B_{c}^{\ast}$ meson is the
   radiative transition process,
   $B_{c}^{\ast}$ ${\to}$ $B_{c}$ $+$ ${\gamma}$.
   This partial decay width can be written as
   \cite{fayyazuddin},
     \begin{equation}
    {\Gamma}(B_{c}^{\ast}{\to}B_{c}{\gamma})\, =\,
     \frac{4}{3}\, {\alpha}_{\rm em}\, k^{3}_{\gamma}\,
    {\vert} {\mu}_{B_{c}^{\ast}B_{c}} {\vert}^{2}
     \label{m1-decay-amplitude},
     \end{equation}
   where the center-of-mass momentum of a photon in the rest
   frame of the
   $B_{c}^{\ast}$ meson and the magnetic dipole
   (M1) momentum are respectively defined as,
     \begin{equation}
     k_{\gamma}\, =\,
     \frac{  m_{B_{c}^{\ast}}^{2}-m_{B_{c}}^{2} }
          { 2\,m_{B_{c}^{\ast}} }
     \, {\approx} \, 56\, {\rm MeV}
     \label{photon-momentum},
     \end{equation}
     \begin{equation}
    {\mu}_{B_{c}^{\ast}B_{c}}
     \, =\, {\langle}B_{c}{\vert}
     \sum\limits_{i=b,c}
     \frac{Q_{i}}{2\,m_{i}}\,{\sigma}_{iz}
    {\vert}B_{c}^{\ast}{\rangle}
     \, =\,
     \frac{1}{6}\,\big(
     \frac{2}{m_{c}}-\frac{1}{m_{b}}\big)
     \label{m1}.
     \end{equation}
   It is clear that kinematically, the parity and angular
   momentum conservation in electromagnetic interactions
   require the orbital angular momentum between the $B_{c}$
   meson and photon to be $L$ $=$ $1$.
   In addition, the photon is very soft, resulting in
   a very small phase space.
   Dynamically, the decay width is proportional to the
   fine-structure constant ${\alpha}_{\rm em} $ of the
   electromagnetic interactions and the module
   square of the magnetic dipole momentum,
   while the magnetic dipole momentum is inversely
   proportional to the mass of heavy quarks.
   The combined effects of both kinematical and dynamical
   factors produce a very narrow decay width,
   ${\Gamma}(B_{c}^{\ast}{\to}B_{c}{\gamma})$ ${\approx}$
   $60$ eV \cite{Eur.Phys.J.C.81.1110}.
   A good approximation is the full decay width of
   the $B_{c}^{\ast}$ meson ${\Gamma}_{B_{c}^{\ast}}$
   ${\approx}$ ${\Gamma}(B_{c}^{\ast}{\to}B_{c}{\gamma})$.
   Experimentally, the electromagnetic process
   $B_{c}^{\ast}$ ${\to}$ $B_{c}{\gamma}$ with
   an occurrence probability of almost $100\%$
   should be easily detected at the $e^{+}e^{-}$
   CEPC and FCC-ee colliders,
   thanks to the excellent photon resolution
   of the electromagnetic calorimeter, and thanks to the
   fine performance in reconstruction technology and method
   of the charged particle tracks.
   Because the masses of $B_{c}^{\ast}$ and $B_{c}$ mesons are
   very close, the detection of the photon from the
   $B_{c}^{\ast}$ ${\to}$
   $B_{c}{\gamma}$ process plays a critical role in
   distinguishing between the $B_{c}^{\ast}$ and $B_{c}$
   mesons.
   The identification of the soft photon will
   be very challenging. What's more, the
   photon from the M1 transition $B_{c}^{\ast}$ ${\to}$
   $B_{c}$ $+$ ${\gamma}$ decay is bound to be seriously
   affected by those from bremsstrahlung radiation and
   chaotic electromagnetic backgrounds, which results
   in identification inefficiencies.
   Besides the electromagnetic interactions, the $B_{c}^{\ast}$
   meson can also decay through the weak interactions in the
   standard model (SM) of elementary particles.
   The narrowness of the full width of the $B_{c}^{\ast}$
   meson affords a great potential
   for experimental investigations on the $B_{c}^{\ast}$
   weak decays.
   The $B_{c}^{\ast}$ weak decays will provide a
   good opportunity and an important and useful alternative
   to find and explore the foreseeable $B_{c}^{\ast}$ mesons
   with some novel ways at the future high-luminosity
   and high-precision experiments.

   Compared with the electromagnetic $B_{c}^{\ast}$ decays, there
   are plenty of $B_{c}^{\ast}$ meson weak decay processes.
   Based on the weak interaction couplings among particles,
   both component quarks of the
   $B_{c}^{\ast}$ meson, the heavy
   bottom and charm quarks can transmit into first
   and second family quarks lighter than themselves.
   These $B_{c}^{\ast}$ weak processes can be
   classified into three types, similar to those for
   the pseudoscalar $B_{c}$ meson,
   (1) the bottom quark decays while the charm quark remain
   quiescent as a spectator;
   (2) the charm quark decays while the bottom quark remain
   inactive as a spectator;
   (3) the bottom and charm quarks annihilate into a virtual
   charged $W$ boson.
   The purely leptonic $B_{c}^{\ast}$ decays belonging to
   type (3) will suffer additional complications from the
   final neutrinos.
   The charged hadrons from $B_{c}^{\ast}$ weak decays,
   such as pions and kaons, are relatively easy to identify
   at sensitive particle detectors.
   In this paper, we will estimate the branching ratio for
   $B_{c}^{{\ast}+}$ ${\to}$ $BP$, $BV$,
   $B^{\ast}P$, ${\psi}P$, ${\eta}_{c}P$,
   ${\eta}_{c}V$ decays arising from external $W$
   emission with the factorization approach,
   where $P$ ($V$) denotes the positively charged pion
   and kaon (${\rho}^{+}$ and $K^{{\ast}+}$).
   Some studies \cite{PhysRevD.95.036024,PhysRevD.95.074032,
   J.Phys.G.45.075005,PhysRevD.105.114015}
   have shown that these processes in question have
   relatively large branching ratios among $B_{c}^{\ast}$
   meson weak decays, and should have the priority to be
   investigated experimentally.
   Here, we hope to provide a feasibility analysis of
   searching for the
   $B_{c}^{\ast}$ meson via some particular
   nonleptonic weak decays in the future high-energy and
   high-luminosity experiments.

   The remainder of this paper is organized as follows.
   The effective Hamiltonian for the nonleptonic
   $B_{c}^{\ast}$ weak decays in question is given
   in Section \ref{sec:hamiltonian}.
   Branching ratios and our comments are presented
   in Section \ref{sec:branch}.
   Section \ref{sec:summary} is devoted to a brief summary.
   The decay amplitudes are listed in Appendix.

   \section{Theoretical framework}
   \label{sec:hamiltonian}
   The effective Hamiltonian for the concerned nonleptonic
   $B_{c}^{\ast}$ decays is written as \cite{RevModPhys.68.1125},
     \begin{eqnarray}
    {\cal H}_{\rm eff} &=&
     \frac{G_{F}}{\sqrt{2}}\, V_{cb}^{\ast}\,V_{uq_{1}}\, \big\{
       C_{1}\,O_{1}^{b}+C_{2}\,O_{2}^{b} \big\}
     \label{hamilton-b} \\ &+&
     \frac{G_{F}}{\sqrt{2}}\, V_{cq_{2}}^{\ast}\,V_{uq_{3}}\, \big\{
       C_{1}\,O_{1}^{c}+C_{2}\,O_{2}^{c} \big\}
     \label{hamilton-c},
     \end{eqnarray}
   where Fermi constant $G_{F}$ ${\approx}$
   $1.166{\times}10^{-5}$ ${\rm GeV}^{-2}$ \cite{pdg2022}
   and Wilson coefficients $C_{1,2}$ are process-independent
   couplings.
   Eq.(\ref{hamilton-b}) and Eq.(\ref{hamilton-c}) correspond
   to type (1) and (2) $B_{c}^{\ast}$ decays, respectively.
   $V_{cq}$ and $V_{uq}$ are the Cabibbo-Kobayashi-Maskawa (CKM)
   matrix elements \cite{PhysRevLett.10.531,PTP.49.652}, and
   their magnitudes have been determined experimentally
   \cite{pdg2022}.
     \begin{eqnarray}
    {\vert} V_{ud} {\vert} &=& 0.97370(14)
     \label{ckm-vud}, \\
    {\vert} V_{us} {\vert} &=& 0.2245(8)
     \label{ckm-vus}, \\
    {\vert} V_{cd} {\vert} &=& 0.221(4)
     \label{ckm-vcd}, \\
    {\vert} V_{cs} {\vert} &=& 0.987(11)
     \label{ckm-vcs}, \\
    {\vert} V_{cb} {\vert} &=& 0.0410(14)
     \label{ckm-vcb}.
     \end{eqnarray}
   The expressions of effective tetra-quark operators are written as,
     \begin{eqnarray}
     O_{1}^{b} &=&
     \big[ \bar{b}_{\alpha}\,{\gamma}^{\mu}\,(1-{\gamma}_{5})\,c_{\alpha} \big]\,
     \big[ \bar{u}_{\beta}\,{\gamma}_{\mu}\,(1-{\gamma}_{5})\,q_{1{\beta}} \big]
     \label{operator-o1-b},  \\
     O_{2}^{b} &=&
     \big[ \bar{b}_{\alpha}\,{\gamma}^{\mu}\,(1-{\gamma}_{5})\,c_{\beta} \big]\,
     \big[ \bar{u}_{\beta}\,{\gamma}_{\mu}\,(1-{\gamma}_{5})\,q_{1{\alpha}} \big]
     \label{operator-o2-b}, \\
     O_{1}^{c} &=&
     \big[ \bar{q}_{2{\alpha}}\,{\gamma}^{\mu}\,(1-{\gamma}_{5})\,c_{\alpha} \big]\,
     \big[ \bar{u}_{\beta}\,{\gamma}_{\mu}\,(1-{\gamma}_{5})\,q_{3{\beta}} \big]
     \label{operator-o1-c},  \\
     O_{2}^{c} &=&
     \big[ \bar{q}_{2{\alpha}}\,{\gamma}^{\mu}\,(1-{\gamma}_{5})\,c_{\beta} \big]\,
     \big[ \bar{u}_{\beta}\,{\gamma}_{\mu}\,(1-{\gamma}_{5})\,q_{3{\alpha}} \big]
     \label{operator-o2-c},
     \end{eqnarray}
    where ${\alpha}$ and ${\beta}$ are the color indices,
    $q_{1,2,3}$ $=$ $d$ and $s$.

    It is clear that the calculation of hadronic matrix
    elements (HMEs) is the remaining central missions
    of decay amplitudes.
    HMEs trap the tetra-quark operators between initial
    and final mesons, and involve the long- and
    short-distance contributions.
    Recently, several attractive phenomenological methods
    based on power counting rules and QCD perturbative
    calculations have been developed to deal with HMEs,
    such as the QCD factorization (QCDF) approach built upon the
    collinear approximation \cite{PhysRevLett.83.1914,
    npb.591.313,npb.606.245,plb.488.46,plb.509.263,
    PhysRevD.64.014036,epjc.36.365,PhysRevD.69.054009,
    npb.774.64,PhysRevD.77.074013,JHEP.2016.09.112},
    and the perturbative QCD (pQCD) approach where contributions
    from the transverse momentum of valence quarks and
    Sudakov factors for hadronic wave functions are
    taken into consideration \cite{PhysRevLett.74.4388,
    plb.348.597,PhysRevD.52.3958,PhysRevD.63.074006,
    PhysRevD.63.054008,PhysRevD.63.074009,plb.555.197}.
    The theme of both the QCDF and pQCD approaches is to
    properly factorize the perturbative and
    nonperturbative contributions contained in HMEs,
    and appropriately evaluate their shares.
    The main objective of this paper is to determine whether
    the $B_{c}^{\ast}$ meson can be explored or not in
    the future experiments, so an approximate estimation
    on branching ratio is completely sufficient.
    In this sense, the naive factorization (NF) approach
    \cite{zpc.34.103} will be used and applied to
    nonleptonic $B_{c}^{\ast}$ meson weak decay in our
    calculation.
    The physics picture of the NF approach is very simple and clear.
    The results from the NF approach can be regarded as the leading
    order approximation of those from the QCDF approach.
    What matters is that the NF approach often has good performance
    for nonleptonic $B$ and $D$ weak decays induced by external
    $W$ emission interactions.
    Using the NF approximation as a working hypothesis, it is
    usually assumed that the final state interactions and
    annihilation contributions might be disregarded.
    The product of quark currents in effective tetra-quark
    operators could be replaced by the product of the
    corresponding hadronic currents formed by the
    physical hadrons involved.
    HMEs of tetra-quark operators are separated into two HMEs
    of hadronic currents, which are further parameterized
    by hadronic decay constants and transition form factors.

    Due to the relatively large mass, the vector mesons
    decay dominantly through the strong and/or
    electromagnetic interactions within the SM.
    The weak decays of the vector mesons are
    rare processes and usually draw less attention.
    Following the conventions in Refs.
    \cite{zpc.29.637,PhysRevD.106.036029,ijmpa.30.1550094,
    jhep.1912.102,epjc.54.107},
    the hadronic decay constants and transition form factors
    are defined by HMEs of color-singlet diquark currents.
     \begin{eqnarray}
     j_{\mu}^{V}\, &=& \bar{q}_{1}\,{\gamma}_{\mu}\,q_{2}
     \label{current-v},
     \\
     j_{\mu}^{A}\, &=& \bar{q}_{1}\,{\gamma}_{\mu}\,{\gamma}_{5}\,q_{2}
     \label{current-a},
     \end{eqnarray}
     \begin{eqnarray}
    {\langle}P(k){\vert}\, j_{\mu}^{V}\, {\vert}0{\rangle}
     & =& 0
     \label{decay-pseudoscalar-v},
     \\
    {\langle}P(k){\vert} \, j_{\mu}^{A}\, {\vert}0{\rangle}
     & =&  -i\,f_{P}\,k_{\mu}
     \label{decay-pseudoscalar-a},
     \\
    {\langle}V(k,{\epsilon}){\vert} \, j_{\mu}^{V}\,  {\vert}0{\rangle}
     & =&  f_{V}\,m_{V}\,{\epsilon}_{\mu}
     \label{decay-vector-v},
     \\
    {\langle}V(k,{\epsilon}){\vert} \, j_{\mu}^{A}\, {\vert}0{\rangle}
     & =&  0
     \label{decay-vector-a},
     \end{eqnarray}
     \begin{equation}
    {\langle}P(p_{2}){\vert}\, j_{\mu}^{V}\,
    {\vert} B_{c}^{\ast}({\epsilon},p_{1}){\rangle} \, =\,
    {\varepsilon}_{{\mu}{\nu}{\alpha}{\beta}}\,
    {\epsilon}^{\nu}\, P^{\alpha}\, q^{\beta}
     \frac{ V^{B_{c}^{\ast}{\to}P}(q^{2}) }
          { m_{B_{c}^{\ast}}+m_{P} }
    \label{fromfactor-v2p-v},
    \end{equation}
     \begin{eqnarray} & &
    {\langle}P(p_{2}){\vert}\, j_{\mu}^{A}\,
    {\vert} B_{c}^{\ast}({\epsilon},p_{1}){\rangle}
    \nonumber \\ &=&
   +i\,2\,m_{B_{c}^{\ast}}\,
     \frac{ {\epsilon}{\cdot}q }{q^{2}}\,
     q_{\mu}\, A_{0}^{B_{c}^{\ast}{\to}P}(q^{2})
   +i\,{\epsilon}_{{\mu}}\, ( m_{B_{c}^{\ast}}+m_{P} )\,
     A_{1}^{B_{c}^{\ast}{\to}P}(q^{2})
     \nonumber \\ & &
   +i\,\frac{ {\epsilon}{\cdot}q }{ m_{B_{c}^{\ast}}+m_{P} }\,
     P_{\mu}\, A_{2}^{B_{c}^{\ast}{\to}P}(q^{2})
   -i\,2\,m_{B_{c}^{\ast}}\,
     \frac{ {\epsilon}{\cdot}q }{q^{2}}\,
     q_{\mu}\, A_{3}^{B_{c}^{\ast}{\to}P}(q^{2})
    \label{fromfactor-v2p-a},
    \end{eqnarray}
     \begin{eqnarray} & &
    {\langle}V({\epsilon}_{2},p_{2}){\vert}\, j_{\mu}^{V}\,
    {\vert}B_{c}^{\ast}({\epsilon}_{1},p_{1}){\rangle} \, =\,
   -({\epsilon}_{1}{\cdot}{\epsilon}_{2}^{\ast})\,
    \big\{ P_{\mu}\,V_{1}^{B_{c}^{\ast}{\to}V}(q^{2})
          -q_{\mu}\,V_{2}^{B_{c}^{\ast}{\to}V}(q^{2}) \big\}
    \nonumber \\ & &
   +\frac{ ({\epsilon}_{1}{\cdot}q)\,
           ({\epsilon}_{2}^{\ast}{\cdot}q) }
         { m_{B_{c}^{\ast}}^{2}-m_{V}^{2} }\,
    \big\{ \big[ P_{\mu} -
    \frac{ m_{B_{c}^{\ast}}^{2}-m_{V}^{2} }{ q^{2} }\,q_{\mu}
    \big]\, V_{3}^{B_{c}^{\ast}{\to}V}(q^{2})
   +\frac{ m_{B_{c}^{\ast}}^{2}-m_{V}^{2} }{ q^{2} }\,
           q_{\mu}\, V_{4}^{B_{c}^{\ast}{\to}V}(q^{2}) \big\}
     \nonumber \\ & &
    -({\epsilon}_{1}{\cdot}q)\,
     {\epsilon}_{2,{\mu}}^{\ast}\,V_{5}^{B_{c}^{\ast}{\to}V}(q^{2})
   +({\epsilon}_{2}^{\ast}{\cdot}q)\,
     {\epsilon}_{1,{\mu}}\,V_{6}^{B_{c}^{\ast}{\to}V}(q^{2})
    \label{fromfactor-v2v-v},
    \end{eqnarray}
     \begin{eqnarray} & &
   {\langle}V({\epsilon}_{2},p_{2}){\vert}\,j_{\mu}^{A}\,
   {\vert}B_{c}^{\ast}({\epsilon}_{1},p_{1}){\rangle}
    \nonumber \\ &=&
   -i\,{\varepsilon}_{{\mu}{\nu}{\alpha}{\beta}}\,
   {\epsilon}_{1}^{\alpha}\,
   {\epsilon}_{2}^{{\ast}{\beta}}\,\big\{ \big[ P^{\nu} -
    \frac{ m_{B_{c}^{\ast}}^{2}-m_{V}^{2} }{ q^{2} }\,q^{\nu}
    \big]\, A_{1}^{B_{c}^{\ast}{\to}V}(q^{2})
   +\frac{ m_{B_{c}^{\ast}}^{2}-m_{V}^{2} }{ q^{2} }\,
    q^{\nu}\, A_{2}^{B_{c}^{\ast}{\to}V}(q^{2}) \big\}
    \nonumber \\ & &
   -\frac{ i\,{\varepsilon}_{{\mu}{\nu}{\alpha}{\beta}}\,
           P^{\alpha}\, q^{\beta} }
         { m_{B_{c}^{\ast}}^{2}-m_{V}^{2} }\, \big\{
    ({\epsilon}_{2}^{\ast}{\cdot}q)\,
   {\epsilon}_{1}^{\nu}\,A_{3}^{B_{c}^{\ast}{\to}V}(q^{2})
  - ({\epsilon}_{1}{\cdot}q)\,
   {\epsilon}_{2}^{{\ast},{\nu}}\,A_{4}^{B_{c}^{\ast}{\to}V}(q^{2}) \big\}
    \label{fromfactor-v2v-a},
    \end{eqnarray}
   where $m_{P}$ ($m_{V}$) and $f_{P}$ ($f_{V}$) are the mass
   and decay constant of final pseudoscalar (vector) mesons,
   respectively;
   ${\epsilon}_{i}$ is the polarization vector;
   $A_{i}$ and $V_{i}$ are mesonic transition form factors;
   the momentum $P$ $=$ $p_{1}$ $+$ $p_{2}$ and
   $q$ $=$ $p_{1}$ $-$ $p_{2}$.
   There are some relationships among form factors,
    \begin{equation}
    A_{0}^{B_{c}^{\ast}{\to}P}(0)\, =\,
    A_{3}^{B_{c}^{\ast}{\to}P}(0)
    \label{v2p-a0-a3},
    \end{equation}
    \begin{equation}
    A_{1}^{B_{c}^{\ast}{\to}V}(0)\, =\,
    A_{2}^{B_{c}^{\ast}{\to}V}(0)
    \label{v2v-a0-a2},
    \end{equation}
    \begin{equation}
    V_{3}^{B_{c}^{\ast}{\to}V}(0)\, =\,
    V_{4}^{B_{c}^{\ast}{\to}V}(0)
    \label{v2v-v3-v4}.
    \end{equation}

   The decay constants and form factors are
   nonperturbative hadronic parameters.
   These parameters are universal, and can be obtained from
   data and some nonperturbative methods.
   The experimental data on the $B_{c}^{\ast}$ meson
   are still unavailable.
   Phenomenologically, form factors at the pole
   $q^{2}$ $=$ $0$ are expressed as the overlap integrals
   of mesonic wave functions \cite{zpc.29.637},
   where the mesonic wave functions are nonperturbative
   but process-independent physical quantities.
   For example, the form factors for the
   $B_{c}^{\ast}$ ${\to}$ $B_{d,s}$ transitions have
   been calculated with the Wirbel-Stech-Bauer model
   in Ref. \cite{PhysRevD.95.074032}.
   Recently, the form factors for the $B_{c}^{\ast}$ ${\to}$
   $J/{\psi}$, $B_{d,s}^{\ast}$ transitions have also
   been investigated with the light front quark model
   (LFQM) in Ref. \cite{jhep.1912.102}.
   Additionally, as noted, the latest decay constants obtained
   with LFQM \cite{PhysRevD.75.034019,PhysRevD.98.114018}
   are generally consistent with the data as well as
   those from lattice QCD simulations and QCD sum
   rule approaches.
   Following the calculation in Ref. \cite{jhep.1912.102},
   we obtain the form factors for the $B_{c}^{\ast}$ ${\to}$
   ${\psi}(2S)$, ${\eta}_{c}(1S)$, ${\eta}_{c}(2S)$,
   $B_{d,s}$ transitions with LFQM.
   The numerical values of the form factors are listed in
   Table \ref{tab:formfactor}.
   The theoretical uncertainties are not particularly
   important or worthy of our attention for the moment.
   After all, the magnitude order estimation of branching
   ratios for the nonleptonic $B_{c}^{\ast}$ weak decays
   is quite sufficient for the present purposes.
  \begin{table}[th]
  \caption{Mesonic decay constants
  \cite{pdg2022,Eur.Phys.J.C.81.1110}
  and form factors at the pole $q^{2}$ $=$ $0$
  \cite{jhep.1912.102}.}
  \label{tab:formfactor}
  \begin{ruledtabular}
  \begin{tabular}{cccc}
     $f_{\pi}$      $=$ $130.2{\pm}1.2$ MeV
   & $f_{K}$        $=$ $155.7{\pm}0.3$ MeV
   & $f_{\rho}$     $=$ $207.7{\pm}1.6$ MeV
   & $f_{K^{\ast}}$ $=$ $202.5^{+6.5}_{-6.7}$ MeV \\ \hline
     $V_{1}^{B_{c}^{\ast}{\to}{\psi}(1S)}$ $=$ $0.56$
   & $V_{2}^{B_{c}^{\ast}{\to}{\psi}(1S)}$ $=$ $0.33$
   & $V_{3,4}^{B_{c}^{\ast}{\to}{\psi}(1S)}$ $=$ $0.20$
   & $V_{5}^{B_{c}^{\ast}{\to}{\psi}(1S)}$ $=$ $1.17$ \\
     $V_{6}^{B_{c}^{\ast}{\to}{\psi}(1S)}$ $=$ $0.65$
   & $A_{1,2}^{B_{c}^{\ast}{\to}{\psi}(1S)}$ $=$ $0.54$
   & $A_{3}^{B_{c}^{\ast}{\to}{\psi}(1S)}$ $=$ $0.13$
   & $A_{4}^{B_{c}^{\ast}{\to}{\psi}(1S)}$ $=$ $0.14$ \\ \hline
     $V_{1}^{B_{c}^{\ast}{\to}{\psi}(2S)}$ $=$ $0.39$
   & $V_{2}^{B_{c}^{\ast}{\to}{\psi}(2S)}$ $=$ $0.32$
   & $V_{3,4}^{B_{c}^{\ast}{\to}{\psi}(2S)}$ $=$ $0.12$
   & $V_{5}^{B_{c}^{\ast}{\to}{\psi}(2S)}$ $=$ $0.79$ \\
     $V_{6}^{B_{c}^{\ast}{\to}{\psi}(2S)}$ $=$ $0.48$
   & $A_{1,2}^{B_{c}^{\ast}{\to}{\psi}(2S)}$ $=$ $0.37$
   & $A_{3}^{B_{c}^{\ast}{\to}{\psi}(2S)}$ $=$ $0.13$
   & $A_{4}^{B_{c}^{\ast}{\to}{\psi}(2S)}$ $=$ $0.07$ \\ \hline
     $V_{1}^{B_{c}^{\ast}{\to}B_{s}^{\ast}}$ $=$ $0.63$
   & $V_{2}^{B_{c}^{\ast}{\to}B_{s}^{\ast}}$ $=$ $1.06$
   & $V_{3,4}^{B_{c}^{\ast}{\to}B_{s}^{\ast}}$ $=$ $0.40$
   & $V_{5}^{B_{c}^{\ast}{\to}B_{s}^{\ast}}$ $=$ $3.52$ \\
     $V_{6}^{B_{c}^{\ast}{\to}B_{s}^{\ast}}$ $=$ $3.02$
   & $A_{1,2}^{B_{c}^{\ast}{\to}B_{s}^{\ast}}$ $=$ $0.53$
   & $A_{3}^{B_{c}^{\ast}{\to}B_{s}^{\ast}}$ $=$ $0.73$
   & $A_{4}^{B_{c}^{\ast}{\to}B_{s}^{\ast}}$ $=$ $0.85$ \\ \hline
     $V_{1}^{B_{c}^{\ast}{\to}B^{\ast}}$ $=$ $0.52$
   & $V_{2}^{B_{c}^{\ast}{\to}B^{\ast}}$ $=$ $1.18$
   & $V_{3.4}^{B_{c}^{\ast}{\to}B^{\ast}}$ $=$ $0.40$
   & $V_{5}^{B_{c}^{\ast}{\to}B^{\ast}}$ $=$ $3.15$ \\
     $V_{6}^{B_{c}^{\ast}{\to}B^{\ast}}$ $=$ $2.66$
   & $A_{1,2}^{B_{c}^{\ast}{\to}B^{\ast}}$ $=$ $0.43$
   & $A_{3}^{B_{c}^{\ast}{\to}B^{\ast}}$ $=$ $0.81$
   & $A_{4}^{B_{c}^{\ast}{\to}B^{\ast}}$ $=$ $0.89$ \\ \hline
     $    V^{B_{c}^{\ast}{\to}{\eta}_{c}(1S)}$ $=$ $0.91$
   & $A_{0}^{B_{c}^{\ast}{\to}{\eta}_{c}(1S)}$ $=$ $0.66$
   & $A_{1}^{B_{c}^{\ast}{\to}{\eta}_{c}(1S)}$ $=$ $0.69$
   & $A_{2}^{B_{c}^{\ast}{\to}{\eta}_{c}(1S)}$ $=$ $0.59$ \\ 
     $    V^{B_{c}^{\ast}{\to}{\eta}_{c}(2S)}$ $=$ $0.59$
   & $A_{0}^{B_{c}^{\ast}{\to}{\eta}_{c}(2S)}$ $=$ $0.43$
   & $A_{1}^{B_{c}^{\ast}{\to}{\eta}_{c}(2S)}$ $=$ $0.41$
   & $A_{2}^{B_{c}^{\ast}{\to}{\eta}_{c}(2S)}$ $=$ $0.51$ \\ \hline
     $    V^{B_{c}^{\ast}{\to}B_{s}}$ $=$ $3.40$
   & $A_{0}^{B_{c}^{\ast}{\to}B_{s}}$ $=$ $0.69$
   & $A_{1}^{B_{c}^{\ast}{\to}B_{s}}$ $=$ $0.75$
   & $A_{2}^{B_{c}^{\ast}{\to}B_{s}}$ $=$ $0.96$ \\ 
     $    V^{B_{c}^{\ast}{\to}B}$ $=$ $3.08$
   & $A_{0}^{B_{c}^{\ast}{\to}B}$ $=$ $0.60$
   & $A_{1}^{B_{c}^{\ast}{\to}B}$ $=$ $0.65$
   & $A_{2}^{B_{c}^{\ast}{\to}B}$ $=$ $0.91$
   \end{tabular}
   \end{ruledtabular}
   \end{table}

   \section{Branching ratio}
   \label{sec:branch}
   The branching ratio of nonleptonic $B_{c}^{\ast}$ decays
   is defined as
     \begin{equation}
    {\cal B}r\, =\,
     \frac{ p_{\rm cm} }{ 24\,{\pi}\,m_{B_{c}^{\ast}}^{2}\,{\Gamma}_{B_{c}^{\ast}} }\,
     \overline{ {\vert}{\cal A}{\vert} }^{2}
    \label{eq:branch}.
    \end{equation}
   where $p_{\rm cm}$ is the center-of-mass momentum of
   final states in the rest frame of the $B_{c}^{\ast}$ meson.
   ${\cal A}$ denotes the decay amplitudes, which are
   collected in the Appendix.
   With the input parameters in Table \ref{tab:formfactor}
   and Table  \ref{tab:meson-mass}, we obtain the
   branching ratios, which are listed in Table \ref{tab:br}.

     \begin{table}[th]
     \caption{Mesonic mass (in the unit of MeV) \cite{pdg2022},
     where their central values will be regarded as the default
     inputs unless otherwise specified.}
     \label{tab:meson-mass}
     \begin{ruledtabular}
     \begin{tabular}{cccc}
     $m_{\pi}$            $=$ $139.57$
   & $m_{\rho}$           $=$ $775.26{\pm}0.25$
   & $m_{B_{d}}$          $=$ $5279.65{\pm}0.12$
   & $m_{B_{d}^{\ast}}$   $=$ $5324.70{\pm}0.21$ \\
     $m_{K}$              $=$ $493.677{\pm}0.016$
   & $m_{K^{\ast}}$       $=$ $891.66{\pm}0.26$
   & $m_{{\eta}_{c}(1S)}$ $=$ $2983.9{\pm}0.5$
   & $m_{{\psi}(1S)}$     $=$ $3096.9$ \\
     $m_{B_{s}}$          $=$ $5366.88{\pm}0.14$
   & $m_{B_{s}^{\ast}}$   $=$ $5415.4^{+1.8}_{-1.5}$
   & $m_{{\eta}_{c}(2S)}$ $=$ $3637.5{\pm}1.1$
   & $m_{{\psi}(2S)}$     $=$ $3686.10{\pm}0.06$
     \end{tabular}
     \end{ruledtabular}
     \end{table}
    {\renewcommand{\baselinestretch}{1.3}
     \begin{table}[th]
     \caption{Branching ratios and event numbers of
      nonleptonic $B_{c}^{\ast}$ weak decays,
      assuming that about $10^{8}$, $10^{9}$ and $3{\times}10^{10}$
      $B_{c}^{\ast}$ mesons will be available at the future CEPC,
      FCC-ee and LHCb@HL-LHC experiments, respectively; where
      ${\lambda}$ ${\approx}$ $0.2$ is the phenomenological
      Wolfenstein parameter.}
     \label{tab:br}
     \begin{ruledtabular}
     \begin{tabular}{c|c|c|c|ccc|ccc}
       decay & CKM
     & \multicolumn{5}{c|}{branching ratio}
     & \multicolumn{3}{c}{event numbers}
       \\ \cline{3-10}
     mode  & factor & unit
     & here & \multicolumn{3}{c|}{previous}
     & CEPC & FCC-ee & LHCb
       \\ \hline
       $B_{s}{\pi}^{+}$
     & $V_{cs}^{\ast}V_{ud}$ ${\sim}$ ${\cal O}(1)$
     & $10^{-7}$
     & $4.4$
     & $4.0$ \cite{PhysRevD.95.074032}
     & $7.3$ \cite{PhysRevD.95.074032}
     & $9.8$ \cite{PhysRevD.95.074032}
     & $44$
     & $442$
     & $13256$
       \\
       $B_{s}{\rho}^{+}$
     & $V_{cs}^{\ast}V_{ud}$ ${\sim}$ ${\cal O}(1)$
     & $10^{-6}$
     & $1.9$
     & $0.6$ \cite{PhysRevD.95.074032}
     & $1.2$ \cite{PhysRevD.95.074032}
     & $1.7$ \cite{PhysRevD.95.074032}
     & $193$
     & $1930$
     & $57887$
       \\
       $B_{s}^{\ast}{\pi}^{+}$
     & $V_{cs}^{\ast}V_{ud}$ ${\sim}$ ${\cal O}(1)$
     & $10^{-6}$
     & $1.4$
     &
     &
     &
     & $143$
     & $1428$
     & $42848$
       \\
       $B_{s}K^{+}$
     & $V_{cs}^{\ast}V_{us}$ ${\sim}$ ${\cal O}({\lambda})$
     & $10^{-8}$
     & $2.2$
     & $2.0$ \cite{PhysRevD.95.074032}
     & $3.6$ \cite{PhysRevD.95.074032}
     & $4.9$ \cite{PhysRevD.95.074032}
     & $2$
     & $22$
     & $658$
       \\
       $B_{s}K^{{\ast}+}$
     & $V_{cs}^{\ast}V_{us}$ ${\sim}$ ${\cal O}({\lambda})$
     & $10^{-8}$
     & $7.2$
     & $2.2$ \cite{PhysRevD.95.074032}
     & $4.1$ \cite{PhysRevD.95.074032}
     & $6.0$ \cite{PhysRevD.95.074032}
     & $7$
     & $72$
     & $2169$
       \\
       $B_{s}^{\ast}K^{+}$
     & $V_{cs}^{\ast}V_{us}$ ${\sim}$ ${\cal O}({\lambda})$
     & $10^{-8}$
     & $8.4$
     &
     &
     &
     & $8$
     & $84$
     & $2533$
       \\
       $B_{d}{\pi}^{+}$
     & $V_{cd}^{\ast}V_{ud}$ ${\sim}$ ${\cal O}({\lambda})$
     & $10^{-8}$
     & $2.1$
     & $2.1$ \cite{PhysRevD.95.074032}
     & $4.5$ \cite{PhysRevD.95.074032}
     & $6.5$ \cite{PhysRevD.95.074032}
     & $2$
     & $21$
     & $640$
       \\
       $B_{d}{\rho}^{+}$
     & $V_{cd}^{\ast}V_{ud}$ ${\sim}$ ${\cal O}({\lambda})$
     & $10^{-8}$
     & $9.7$
     & $3.5$ \cite{PhysRevD.95.074032}
     & $7.5$ \cite{PhysRevD.95.074032}
     & $11.5$ \cite{PhysRevD.95.074032}
     & $10$
     & $97$
     & $2924$
       \\
       $B_{d}^{\ast}{\pi}^{+}$
     & $V_{cd}^{\ast}V_{ud}$ ${\sim}$ ${\cal O}({\lambda})$
     & $10^{-8}$
     & $6.3$
     &
     &
     &
     & $6$
     & $63$
     & $1884$
       \\
       $B_{d}K^{+}$
     & $V_{cd}^{\ast}V_{us}$ ${\sim}$ ${\cal O}({\lambda}^{2})$
     & $10^{-9}$
     & $1.1$
     & $1.2$ \cite{PhysRevD.95.074032}
     & $2.4$ \cite{PhysRevD.95.074032}
     & $3.5$ \cite{PhysRevD.95.074032}
     & $0$
     & $1$
     & $34$
       \\
       $B_{d}K^{{\ast}+}$
     & $V_{cd}^{\ast}V_{us}$ ${\sim}$ ${\cal O}({\lambda}^{2})$
     & $10^{-9}$
     & $4.4$
     & $1.5$ \cite{PhysRevD.95.074032}
     & $3.1$ \cite{PhysRevD.95.074032}
     & $4.8$ \cite{PhysRevD.95.074032}
     & $0$
     & $4$
     & $133$
       \\
       $B_{d}^{\ast}K^{+}$
     & $V_{cd}^{\ast}V_{us}$ ${\sim}$ ${\cal O}({\lambda}^{2})$
     & $10^{-9}$
     & $3.9$
     &
     &
     &
     & $0$
     & $4$
     & $117$
       \\ \hline
       ${\eta}_{c}(1S){\pi}^{+}$
     & $V_{cb}^{\ast}V_{ud}$ ${\sim}$ ${\cal O}({\lambda}^{2})$
     & $10^{-8}$
     & $1.5$
     & $2.2$ \cite{PhysRevD.95.036024}
     &
     &
     & $2$
     & $15$
     & $455$
       \\
       ${\eta}_{c}(1S){\rho}^{+}$
     & $V_{cb}^{\ast}V_{ud}$ ${\sim}$ ${\cal O}({\lambda}^{2})$
     & $10^{-8}$
     & $4.3$
     & $3.0$ \cite{J.Phys.G.45.075005}
     & $2.5$ \cite{J.Phys.G.45.075005}
     &
     & $4$
     & $43$
     & $1300$
       \\
       ${\psi}(1S){\pi}^{+}$
     & $V_{cb}^{\ast}V_{ud}$ ${\sim}$ ${\cal O}({\lambda}^{2})$
     & $10^{-8}$
     & $5.1$
     & $9.2$ \cite{PhysRevD.95.036024}
     & $2.4$ \cite{PhysRevD.105.114015}
     &
     & $5$
     & $51$
     & $1519$
       \\
       ${\eta}_{c}(2S){\pi}^{+}$
     & $V_{cb}^{\ast}V_{ud}$ ${\sim}$ ${\cal O}({\lambda}^{2})$
     & $10^{-9}$
     & $4.1$
     & $2.4$ \cite{PhysRevD.95.036024}
     &
     &
     & $0$
     & $4$
     & $123$
       \\
       ${\eta}_{c}(2S){\rho}^{+}$
     & $V_{cb}^{\ast}V_{ud}$ ${\sim}$ ${\cal O}({\lambda}^{2})$
     & $10^{-8}$
     & $1.2$
     &
     &
     &
     & $1$
     & $12$
     & $351$
       \\
       ${\psi}(2S){\pi}^{+}$
     & $V_{cb}^{\ast}V_{ud}$ ${\sim}$ ${\cal O}({\lambda}^{2})$
     & $10^{-8}$
     & $1.6$
     & $3.2$ \cite{PhysRevD.95.036024}
     &
     &
     & $2$
     & $16$
     & $471$
       \\
       ${\eta}_{c}(1S)K^{+}$
     & $V_{cb}^{\ast}V_{us}$ ${\sim}$ ${\cal O}({\lambda}^{3})$
     & $10^{-9}$
     & $1.1$
     & $1.7$ \cite{PhysRevD.95.036024}
     &
     &
     & $0$
     & $1$
     & $33$
       \\
       ${\eta}_{c}(1S)K^{{\ast}+}$
     & $V_{cb}^{\ast}V_{us}$ ${\sim}$ ${\cal O}({\lambda}^{3})$
     & $10^{-9}$
     & $2.3$
     & $1.7$ \cite{J.Phys.G.45.075005}
     & $1.4$ \cite{J.Phys.G.45.075005}
     &
     & $0$
     & $2$
     & $68$
       \\
       ${\psi}(1S)K^{+}$
     & $V_{cb}^{\ast}V_{us}$ ${\sim}$ ${\cal O}({\lambda}^{3})$
     & $10^{-9}$
     & $3.8$
     & $7.3$ \cite{PhysRevD.95.036024}
     &
     &
     & $0$
     & $4$
     & $113$
       \\
       ${\eta}_{c}(2S)K^{+}$
     & $V_{cb}^{\ast}V_{us}$ ${\sim}$ ${\cal O}({\lambda}^{3})$
     & $10^{-10}$
     & $2.9$
     & $3.4$ \cite{PhysRevD.95.036024}
     &
     &
     & $0$
     & $0$
     & $9$
       \\
       ${\eta}_{c}(2S)K^{{\ast}+}$
     & $V_{cb}^{\ast}V_{us}$ ${\sim}$ ${\cal O}({\lambda}^{3})$
     & $10^{-10}$
     & $6.1$
     &
     &
     &
     & $0$
     & $1$
     & $18$
       \\
       ${\psi}(2S)K^{+}$
     & $V_{cb}^{\ast}V_{us}$ ${\sim}$ ${\cal O}({\lambda}^{3})$
     & $10^{-9}$
     & $1.2$
     & $2.4$ \cite{PhysRevD.95.036024}
     &
     &
     & $0$
     & $1$
     & $35$
     \end{tabular}
     \end{ruledtabular}
     \end{table} }

   Our comments on branching ratios are as follows.

   (1)
   For the nonleptonic $B_{c}^{\ast}$ weak decays induced by
   the external $W$ emission interactions,
   branching ratios using the NF approach are generally
   of the same
   order of magnitude as previous estimations with the
   QCD-inspired QCDF and pQCD approach
   \cite{PhysRevD.95.036024,PhysRevD.95.074032,
   J.Phys.G.45.075005}.
   The consistency of the results with different approaches
   indicates that our results in this paper might be reasonable.
   Besides, the branching ratios for the $B_{c}^{\ast}$ ${\to}$
   $B_{s,d}^{\ast}{\pi}$, $B_{s,d}^{\ast}K$,
   ${\eta}_{c}(2S){\rho}$, ${\eta}_{c}(2S)K^{\ast}$
   decays are estimated for the first time.
   It is found that the $B_{c}^{\ast}$ ${\to}$ $B_{s}^{\ast}{\pi}$
   decay has a relatively large branching ratio among the
   two-meson $B_{c}^{\ast}$ weak decays.
    In addition, both the initial $B_{c}^{\ast}$ meson
    and one of the final meson of the processes in question
    contain one and/or two of the heavy quarks.
    Due to the fact that the light quarks and gluon clouds
    are almost blind to the spin of the heavy quark with
    the approximation of the heavy quark limit,
    the heavy quark spin symmetry should be expected to
    relate the initial and final mesons \cite{Z.Phys.C.62.271}.
    In the extreme non-relativistic limit, it is expected
    to have \cite{Z.Phys.C.62.271}
     \begin{equation}
     r\, =\,
     \frac{ {\cal B}r( 1^{-} {\to} 1^{-} {\pi} ) }
          { {\cal B}r( 1^{-} {\to} 0^{-} {\pi} ) }\
    {\approx}\
     \frac{\displaystyle \left.
     \frac{ d{\cal B}r( 1^{-} {\to} 1^{-} {\ell}{\nu} ) }
          { dq^{2} } \right\vert_{q^{2}\, =\, m_{\pi}^{2} } }
          {\displaystyle \left.
     \frac{ {\cal B}r( 1^{-} {\to} 0^{-} {\ell}{\nu} ) }
          { dq^{2} } \right\vert_{q^{2}\, =\, m_{\pi}^{2} } }\
    {\approx}\ 3
     \label{r-br-heavy-quark-spin}.
     \end{equation}
    The mass difference in the phase factors will
    furnish the above relations in Eq.(\ref{r-br-heavy-quark-spin}).
    Here, some relative branching ratios are,
     \begin{equation}
     r_{1}\, =\,
     \frac{ {\cal B}r( B_{c}^{\ast} {\to} B_{s}^{\ast} {\pi} ) }
          { {\cal B}r( B_{c}^{\ast} {\to} B_{s}        {\pi} ) }
     \, =\, 3.2
     \label{r-br-bs-pi},
     \end{equation}
     \begin{equation}
     r_{2}\, =\,
     \frac{ {\cal B}r( B_{c}^{\ast} {\to} B_{d}^{\ast} {\pi} ) }
          { {\cal B}r( B_{c}^{\ast} {\to} B_{d}        {\pi} ) }
     \, =\, 2.9
     \label{r-br-bd-pi},
     \end{equation}
     \begin{equation}
     r_{3}\, =\,
     \frac{ {\cal B}r( B_{c}^{\ast} {\to} B_{s}^{\ast} K ) }
          { {\cal B}r( B_{c}^{\ast} {\to} B_{s}        K ) }
     \, =\, 3.8
     \label{r-br-bs-k},
     \end{equation}
     \begin{equation}
     r_{4}\, =\,
     \frac{ {\cal B}r( B_{c}^{\ast} {\to} B_{d}^{\ast} K ) }
          { {\cal B}r( B_{c}^{\ast} {\to} B_{d}        K ) }
     \, =\, 3.4
     \label{r-br-bd-k},
     \end{equation}
     \begin{equation}
     r_{5}\, =\,
     \frac{ {\cal B}r( B_{c}^{\ast} {\to} {\psi}(1S) {\pi} ) }
          { {\cal B}r( B_{c}^{\ast} {\to} {\eta}_{c}(1S) {\pi} ) }
     \, =\, 3.3
     \label{r-br-cc1s-pi},
     \end{equation}
     \begin{equation}
     r_{6}\, =\,
     \frac{ {\cal B}r( B_{c}^{\ast} {\to} {\psi}(1S) K ) }
          { {\cal B}r( B_{c}^{\ast} {\to} {\eta}_{c}(1S) K ) }
     \, =\, 3.4
     \label{r-br-cc1s-k},
     \end{equation}
     \begin{equation}
     r_{7}\, =\,
     \frac{ {\cal B}r( B_{c}^{\ast} {\to} {\psi}(2S) {\pi} ) }
          { {\cal B}r( B_{c}^{\ast} {\to} {\eta}_{c}(2S) {\pi} ) }
     \, =\, 3.8
     \label{r-br-cc2s-pi},
     \end{equation}
     \begin{equation}
     r_{8}\, =\,
     \frac{ {\cal B}r( B_{c}^{\ast} {\to} {\psi}(2S) K ) }
          { {\cal B}r( B_{c}^{\ast} {\to} {\eta}_{c}(2S) K ) }
     \, =\, 3.9
     \label{r-br-cc2s-k}.
     \end{equation}
    For the charm quark decay, the ratio
    $r_{1}$ of the Cabibbo-favored $B_{c}^{\ast}$ ${\to}$
    $B_{s}^{\ast}{\pi}$, $B_{s}{\pi}$ decays
    is generally consistent with both the expectation
    Eq.(\ref{r-br-heavy-quark-spin}) from the heavy
    quark spin symmetry
    and the ratio of semileptonic $J/{\psi}$ weak decays
    ${\cal B}r( J/{\psi} {\to} D_{s}^{\ast}e^{+}{\nu}_{e} )/
     {\cal B}r( J/{\psi} {\to} D_{s}  e^{+}{\nu}_{e} )$
    ${\approx}$ $3.1$ obtained with the QCD sum rules
    \cite{epjc.54.107}.

   (2)
   For the $B_{c}^{\ast}$ decays induced either
   by the bottom quark decay [type (1)] or by the charm quark decay
   [type (2)], there are some clear hierarchical relationship
   among branching ratios according to the CKM factors of
   decay amplitudes.
   The Cabibbo-favored $B_{c}^{\ast}$ ${\to}$ 
   $B_{s}^{\ast}{\pi}$ and $B_{s}{\rho}$ decays have
   relatively maximum branching ratios which can reach up
   to ${\cal O}(10^{-6})$.
   The CKM-suppressed $B_{c}^{\ast}$ ${\to}$ ${\eta}_{c}(1S,2S)K$,
   ${\psi}(1S,2S)K$, and ${\eta}_{c}(1S,2S)K^{\ast}$ decays, whose
   amplitudes are proportional to Wolfenstein parameter
   ${\lambda}^{3}$, have relatively minimal branching ratios.
   In addition, there are three partial wave amplitudes for
   final states containing one pseudoscalar plus vector mesons,
   while there is only the $p$-wave amplitude for two final
   pseudoscalar mesons. Hence, there are some relationships,
   ${\cal B}r(B_{c}^{\ast}{\to}B_{s}^{\ast}{\pi})$ $>$
   ${\cal B}r(B_{c}^{\ast}{\to}B_{s}{\pi})$ and
   ${\cal B}r(B_{c}^{\ast}{\to}B_{s}{\rho})$ $>$
   ${\cal B}r(B_{c}^{\ast}{\to}B_{s}{\pi})$, and
   these hierarchical relationships are also true for
   other similar cases where final states have the
   same valence quark compositions but different
   orbit-spin couplings among quarks.

   (3)
   A more careful theoretical investigation of these decays
   in question is necessary and helpful to explore the
   hadronic $B_{c}^{\ast}$ weak decays experimentally.
   It should be pointed out that many influences can affect
   the final numerical results on nonleptonic $B_{c}^{\ast}$
   weak decays, such as the nonfactorizable contributions.
   It is clear from Eq.(\ref{ckm-vud}-\ref{ckm-vcb}) that
   ${\vert}V_{cb}{\vert}$ has the largest uncertainty,
   about $3\%$, among the CKM elements involved.
   Except for the decay constant of $f_{K^{\ast}}$,
   the uncertainties from other decay constants in
   Table \ref{tab:formfactor} is less than $1\%$.
   For the nonleptonic decays induced by the external $W$
   boson emission interactions,
   the nonfactorizable contributions to the coefficient
   $a_{1}$ is about $15\%$ for charm quark decays
   \cite{ijmpa.30.1550094}, {\it i.e.},
   the $B_{c}^{\ast}$ ${\to}$ $B_{d,s}^{(\ast)}$
   transitions,
   and about $5\%$ for bottom quark decays
   \cite{PhysRevD.77.074013,JHEP.2016.09.112},
   {\it i.e.},
   the $B_{c}^{\ast}$ ${\to}$ ${\psi}$ and ${\eta}_{c}$
   transitions.
   Large theoretical uncertainties come mainly from the hadronic
   transition form factors, and some of them can reach
   about $40\%$ \cite{jhep.1912.102}.
   In addition, the undetermined decay width of
   the $B_{c}^{\ast}$ meson (for example,
   ${\Gamma}(B_{c}^{\ast}{\to}{\gamma}B_{c})$
   $=$ $20$ ${\sim}$ $135$ eV,
   see Table 2 of
   Ref. \cite{J.Phys.G.45.075005}) will bring very large
   theoretical uncertainties to branching ratios.
   An accurate theoretical prediction seems to be temporarily unavailable.
   All numbers in Table \ref{tab:br} are rough estimates
   and only indicative of the expected order of magnitude.
   Here, what matters to us is whether it is possible to study
   the unacquainted $B_{c}^{\ast}$ mesons in future
   experiments. Hence, a rough estimate is sufficient.

   (4)
   As it is well known that the vector ${\rho}$ and
   $K^{\ast}$ mesons are resonances,  they will
   decay immediately via the strong interactions,
   with branching ratios
   ${\cal B}r({\rho}{\to}{\pi}{\pi})$ ${\sim}$ $100\,\%$
   and ${\cal B}r(K^{\ast}{\to}K{\pi})$ ${\sim}$ $100\,\%$
   \cite{pdg2022}.
   The vector ${\rho}$ ($K^{\ast}$) meson is
   reconstructed experimentally by the final
   pseudoscalar mesons.
   An educated guess is that the branching ratios for
   the three-body decay modes,
   $B_{c}^{\ast}$ ${\to}$ $B_{s,d}{\pi}{\pi}$,
   $B_{s,d}{\pi}K$, ${\eta}_{c}{\pi}{\pi}$,
   ${\eta}_{c}{\pi}K$, should be of a similar order of
   magnitude to those for the $B_{c}^{\ast}$
   ${\to}$ $B_{s,d}{\rho}$, $B_{s,d}K^{\ast}$,
   ${\eta}_{c}{\rho}$, ${\eta}_{c}K^{\ast}$,
   decays, respectively.
   If the contributions from other possible resonances
   reconstructed from the ${\pi}{\pi}$, ${\pi}K$,
   $B_{s,d}{\pi}$, $B_{s,d}K$, ${\eta}_{c}{\pi}$
   and ${\eta}_{c}K$ states
   are taken into consideration,
   the above hadronic three-body $B_{c}^{\ast}$ decays
   are likely to have  even larger branching ratios.
   All in all,
   it is not utopian to expected that the Cabibbo-favored
   $B_{c}^{\ast}$ ${\to}$ $B_{s}{\pi}$, $B_{s}^{\ast}{\pi}$,
   $B_{s}{\rho}$ decays, the singly-Cabibbo-suppressed
   $B_{c}^{\ast}$ ${\to}$ $B_{s}K$, $B_{s}^{\ast}K$,
   $B_{s}K^{\ast}$, $B_{d}{\pi}$, $B_{d}^{\ast}{\pi}$,
   $B_{d}{\rho}$ decays, and even the CKM-suppressed
   $B_{c}^{\ast}$ ${\to}$ ${\eta}_{c}(1S){\pi}$,
   ${\eta}_{c}(1S,2S){\rho}$, ${\psi}(1S,2S){\pi}$
   decays might be observable at the future CEPC,
   FCC-ee and LHCb@HL-LHC experiments.

  \section{Summary}
  \label{sec:summary}
   It has been established that the charged ground vector $B_{c}^{\ast}$
   meson carrying explicit flavor numbers should really exist
   according to the quark model, but to date this has been
   merely on the theoretical
   calculation level rather than the experimental measurement
   level.
   The identification of the
   $B_{c}^{\ast}$ meson at experiments
   is necessary and of important significance to
   the quark model and SM.
   The signal reconstruction of the
   $B_{c}^{\ast}$ meson from the  electromagnetic decay
   $B_{c}^{\ast}$ ${\to}$ $B_{c}{\gamma}$
   is severely hindered by the electromagnetic background
   pollution.
   Inspired by the prospects of huge numbers of
   the $B_{c}^{\ast}$
   mesons in future high-energy and high-luminosity colliders,
   an attractive and competitive choice might be searching for
   the $B_{c}^{\ast}$ meson from its nonleptonic weak decays, where
   the charged final hadrons are comparatively easily and
   effectively  identified in experiments.
   In this paper, we study two kinds of nonleptonic $B_{c}^{\ast}$
   meson weak decays resulting from external $W$ boson emission
   interactions, by using the factorization approximation and
   form factors from light front quark model, one
   originating from bottom quark decays, and the other from
   charm quark decays.
   It is found that the branching ratios for the Cabibbo-favored
   $B_{c}^{\ast}$ ${\to}$ $B_{s}^{\ast}{\pi}$, $B_{s}{\rho}$
   decays can reach up to ${\cal O}(10^{-6})$, and have the
   first priority to be studied at experiments.
   For the singly-Cabibbo-suppressed
   $B_{c}^{\ast}$ ${\to}$ $B_{s}K$, $B_{s}^{\ast}K$,
   $B_{s}K^{\ast}$, $B_{d}{\pi}$, $B_{d}^{\ast}{\pi}$,
   $B_{d}{\rho}$ decays and the CKM-suppressed
   $B_{c}^{\ast}$ ${\to}$ ${\eta}_{c}(1S,2S){\pi}$,
   ${\eta}_{c}(1S,2S){\rho}$, ${\psi}(1S,2S){\pi}$
   decays, several hundred or even thousands of
   events might be observable at CEPC, FCC-ee and
   LHCb@HL-LHC experiments.
   This study provides a ready and helpful reference for
   experimental discovery and investigation of $B_{c}^{\ast}$
   mesons in the future.

  \section*{Acknowledgments}
  The work is supported by the National Natural Science Foundation
  of China (Grant Nos. 11705047, 12275068, U1632109, 11875122)
  and Natural Science Foundation of Henan Province
  (Grant No. 222300420479),
  the Excellent Youth Foundation of Henan Province
  (Grant No. 212300410010).

  \begin{appendix}

   \section{amplitudes for nonleptonic $B_{c}^{\ast}$ weak decays}
   \label{blocks}
   With the conventions of Ref. \cite{PhysRevD.106.036029},
   the amplitudes for nonleptonic $B_{c}^{\ast}$ weak
   decays can be written as follows.

   Based on the conservation of angular momentum,
   there are only $p$-wave amplitudes contributing to
   $B_{c}^{\ast}$ meson decay into two pseudoscalar mesons.
     \begin{eqnarray}
    {\cal A}(B_{c}^{{\ast}+}{\to}B_{s}^{0}{\pi}^{+})
     & = &
     V_{cs}^{\ast}\,V_{ud}\, f_{\pi}\, {\cal M}_{p}^{B_{s}}\,
     ({\epsilon}_{B_{c}^{\ast}}{\cdot}p_{B_{s}})
     \label{bs-pi}, \\
    {\cal A}(B_{c}^{{\ast}+}{\to}B_{s}^{0}K^{+})
     & = &
     V_{cs}^{\ast}\,V_{us}\, f_{K}\, {\cal M}_{p}^{B_{s}}\,
     ({\epsilon}_{B_{c}^{\ast}}{\cdot}p_{B_{s}})
     \label{bs-k}, \\
    {\cal A}(B_{c}^{{\ast}+}{\to}B_{d}^{0}{\pi}^{+})
     & = &
     V_{cd}^{\ast}\,V_{ud}\, f_{\pi}\,  {\cal M}_{p}^{B}\,
     ({\epsilon}_{B_{c}^{\ast}}{\cdot}p_{B_{d}})
     \label{bd-pi}, \\
    {\cal A}(B_{c}^{{\ast}+}{\to}B_{d}^{0}K^{+})
     & = &
     V_{cd}^{\ast}\,V_{us}\, f_{K}\, {\cal M}_{p}^{B}\,
     ({\epsilon}_{B_{c}^{\ast}}{\cdot}p_{B_{d}})
     \label{bd-k}, \\
    {\cal A}(B_{c}^{{\ast}+}{\to}{\eta}_{c}{\pi}^{+})
     & = &
     V_{cb}^{\ast}\,V_{ud}\, f_{\pi}\, {\cal M}_{p}^{{\eta}_{c}}\,
     ({\epsilon}_{B_{c}^{\ast}}{\cdot}p_{{\eta}_{c}})
     \label{etac-pi}, \\
    {\cal A}(B_{c}^{{\ast}+}{\to}{\eta}_{c}K^{+})
     & = &
     V_{cd}^{\ast}\,V_{us}\, f_{K}\, {\cal M}_{p}^{{\eta}_{c}}\,
     ({\epsilon}_{B_{c}^{\ast}}{\cdot}p_{{\eta}_{c}})
     \label{etac-k},
     \end{eqnarray}
   with the common factor of $p$-wave partial amplitudes,
     \begin{eqnarray}
    {\cal M}_{p}^{B_{s}}  & = &
     \sqrt{2}\, G_{F}\, a_{1}\, m_{B_{c}^{\ast}}\,
      A_{0}^{B_{c}^{\ast}{\to}B_{s}}
     \label{amp-p-bs}, \\
    {\cal M}_{p}^{B}  & = &
     \sqrt{2}\, G_{F}\,  a_{1} \, m_{B_{c}^{\ast}}\,
      A_{0}^{B_{c}^{\ast}{\to}B}
     \label{amp-p-bd}, \\
    {\cal M}_{p}^{{\eta}_{c}}  & = &
     \sqrt{2}\, G_{F}\,  a_{1} \, m_{B_{c}^{\ast}}\,
      A_{0}^{B_{c}^{\ast}{\to}{\eta}_{c}}
     \label{amp-p-etac},
     \end{eqnarray}
   where coefficient $a_{1}$ $=$ $C_{1}$ $+$ $C_{2}/N_{c}$
   is generally influenced by nonfactorizable contributions,
   and can be regarded as a phenomenological parameter, especially
   for charm quark decays.
   The value of $a_{1}$ ${\approx}$ $1.2$ will be used in
   our calculation.

   There are three partial wave amplitudes contributing to
   $B_{c}^{\ast}$ meson decay into one pseudoscalar meson
   plus one vector meson.
   The general decay amplitude is written as,
     \begin{eqnarray}
    {\cal A}(B_{c}^{\ast}{\to}VP) &=&
    {\cal M}_{s}\,
     ( {\epsilon}_{B_{c}^{\ast}}{\cdot}{\epsilon}_{V}^{\ast} )
     \nonumber \\ &+&
     \frac{ {\cal M}_{d} }{ m_{V_{B_{c}^{\ast}}}\,m_{V} }\,
     ( {\epsilon}_{B_{c}^{\ast}}{\cdot}p_{V} ) \,
     ( {\epsilon}_{V}^{\ast}{\cdot}p_{B_{c}^{\ast}} )
     \nonumber \\ & +&
     \frac{ {\cal M}_{p} }{ m_{B_{c}^{\ast}}\,m_{V} }\,
     {\varepsilon}_{{\mu}{\nu}{\alpha}{\beta}}\,
     {\epsilon}_{B_{c}^{\ast}}^{\mu}\,
     {\epsilon}_{V}^{{\ast}{\nu}}\,
      p_{B_{c}^{\ast}}^{\alpha}\, p_{V}^{\beta}
     \label{vp-amplitude},
     \end{eqnarray}
   where ${\cal M}_{s,p,d}$ correspond to the $s$-, $p$-,
   and $d$-wave partial amplitudes.

   For $B_{c}^{{\ast}+}$ ${\to}$ $B_{s}^{{\ast}0}{\pi}^{+}$,
   $B_{s}^{{\ast}0}K^{+}$ decays, one has
     \begin{eqnarray}
    {\cal M}_{s}^{ B_{s}^{\ast}{\pi} } &=&
     -i\, \frac{ G_{F} }{ \sqrt{2} }\, V_{cs}^{\ast}\,V_{ud}\,
     f_{ \pi }\, a_{1}\,
     \big( m_{B_{c}^{\ast}}^{2}-m_{B_{s}^{\ast}}^{2} \big)\,
     V_{1}^{ B_{c}^{\ast} {\to} B_{s}^{\ast} }
     \label{bsv-pi-s}, \\
    {\cal M}_{d}^{ B_{s}^{\ast}{\pi} } &=&
     -i\, \frac{ G_{F} }{ \sqrt{2} }\, V_{cs}^{\ast}\,V_{ud}\,
     f_{ \pi }\, a_{1}\, m_{B_{c}^{\ast}}\, m_{B_{s}^{\ast}}\,
     \big( V_{4}^{ B_{c}^{\ast} {\to} B_{s}^{\ast} }
         - V_{5}^{ B_{c}^{\ast} {\to} B_{s}^{\ast} }
         + V_{6}^{ B_{c}^{\ast} {\to} B_{s}^{\ast} } \big)
     \label{bsv-pi-d}, \\
    {\cal M}_{p}^{ B_{s}^{\ast}{\pi} } &=&
     -2\, \frac{ G_{F} }{ \sqrt{2} }\, V_{cs}^{\ast}\,V_{ud}\,
     f_{ \pi }\, a_{1}\, m_{B_{c}^{\ast}}\, m_{B_{s}^{\ast}}\,
       A_{1}^{ B_{c}^{\ast} {\to} B_{s}^{\ast} }
     \label{bsv-pi-p}, \\
    {\cal M}_{i}^{ B_{s}^{\ast}K } &=&
    {\cal M}_{i}^{ B_{s}^{\ast}{\pi} }
     \big( V_{ud}{\to}V_{us},\,f_{\pi}{\to}f_{K} \big),
     \quad \text{for } i=s,p,d
     \label{bsv-k}.
     \end{eqnarray}

   For $B_{c}^{{\ast}+}$ ${\to}$ $B_{d}^{{\ast}0}{\pi}^{+}$,
   $B_{d}^{{\ast}0}K^{+}$ decays, one has
     \begin{eqnarray}
    {\cal M}_{s}^{ B_{d}^{\ast}{\pi} } &=&
     -i\, \frac{ G_{F} }{ \sqrt{2} }\, V_{cd}^{\ast}\,V_{ud}\,
     f_{ \pi }\, a_{1}\,
     \big( m_{B_{c}^{\ast}}^{2}-m_{B_{d}^{\ast}}^{2} \big)\,
     V_{1}^{ B_{c}^{\ast} {\to} B_{d}^{\ast} }
     \label{bdv-pi-s}, \\
    {\cal M}_{d}^{ B_{d}^{\ast}{\pi} } &=&
     -i\, \frac{ G_{F} }{ \sqrt{2} }\, V_{cd}^{\ast}\,V_{ud}\,
     f_{ \pi }\, a_{1}\, m_{B_{c}^{\ast}}\, m_{B_{d}^{\ast}}\,
     \big( V_{4}^{ B_{c}^{\ast} {\to} B_{d}^{\ast} }
         - V_{5}^{ B_{c}^{\ast} {\to} B_{d}^{\ast} }
         + V_{6}^{ B_{c}^{\ast} {\to} B_{d}^{\ast} } \big)
     \label{bdv-pi-d}, \\
    {\cal M}_{p}^{ B_{d}^{\ast}{\pi} } &=&
     -2\, \frac{ G_{F} }{ \sqrt{2} }\, V_{cd}^{\ast}\,V_{ud}\,
     f_{ \pi }\, a_{1}\, m_{B_{c}^{\ast}}\, m_{B_{d}^{\ast}}\,
       A_{1}^{ B_{c}^{\ast} {\to} B_{d}^{\ast} }
     \label{bdv-pi-p}, \\
    {\cal M}_{i}^{ B_{d}^{\ast}K } &=&
    {\cal M}_{i}^{ B_{d}^{\ast}{\pi} }
     \big( V_{ud}{\to}V_{us},\,f_{\pi}{\to}f_{K} \big),
     \quad \text{for } i=s,p,d
     \label{bdv-k}.
     \end{eqnarray}

   For $B_{c}^{{\ast}+}$ ${\to}$ ${\psi}{\pi}^{+}$,
   ${\psi}K^{+}$ decays, one has
     \begin{eqnarray}
    {\cal M}_{s}^{ {\psi}{\pi} } &=&
     -i\, \frac{ G_{F} }{ \sqrt{2} }\, V_{cb}^{\ast}\,V_{ud}\,
     f_{ \pi }\, a_{1}\,
     \big( m_{B_{c}^{\ast}}^{2}-m_{\psi}^{2} \big)\,
     V_{1}^{ B_{c}^{\ast} {\to} {\psi} }
     \label{psi-pi-s}, \\
    {\cal M}_{d}^{ {\psi}{\pi} } &=&
     -i\, \frac{ G_{F} }{ \sqrt{2} }\, V_{cb}^{\ast}\,V_{ud}\,
     f_{ \pi }\, a_{1}\, m_{B_{c}^{\ast}}\, m_{\psi}\,
     \big( V_{4}^{ B_{c}^{\ast} {\to} {\psi} }
         - V_{5}^{ B_{c}^{\ast} {\to} {\psi} }
         + V_{6}^{ B_{c}^{\ast} {\to} {\psi} } \big)
     \label{psi-pi-d}, \\
    {\cal M}_{p}^{ {\psi}{\pi} } &=&
     -2\, \frac{ G_{F} }{ \sqrt{2} }\, V_{cb}^{\ast}\,V_{ud}\,
     f_{ \pi }\, a_{1}\, m_{B_{c}^{\ast}}\, m_{\psi}\,
       A_{1}^{ B_{c}^{\ast} {\to} {\psi} }
     \label{psi-pi-p}, \\
    {\cal M}_{i}^{ {\psi}K } &=&
    {\cal M}_{i}^{ {\psi}{\pi} }
     \big( V_{ud}{\to}V_{us},\,f_{\pi}{\to}f_{K} \big),
     \quad \text{for } i=s,p,d
     \label{psi-k}.
     \end{eqnarray}

   For $B_{c}^{{\ast}+}$ ${\to}$ $B_{s}^{0}{\rho}^{+}$,
   $B_{s}^{0}K^{{\ast}+}$ decays, one has
     \begin{eqnarray}
    {\cal M}_{s}^{ B_{s}{\rho} } &=&
     -i\, \frac{ G_{F} }{ \sqrt{2} }\, V_{cs}^{\ast}\,V_{ud}\,
     f_{\rho}\, m_{\rho}\, a_{1}\,
     \big( m_{B_{c}^{\ast}}+m_{B_{s}} \big)\,
     A_{1}^{ B_{c}^{\ast} {\to} B_{s} }
     \label{bs-rho-s}, \\
    {\cal M}_{d}^{ B_{s}{\rho} } &=&
     -i\, \frac{ G_{F} }{ \sqrt{2} }\, V_{cs}^{\ast}\,V_{ud}\,
     f_{\rho}\, m_{\rho}\, a_{1}\,
     \frac{ 2\,m_{B_{c}^{\ast}}\,m_{\rho} }
          { m_{B_{c}^{\ast}}+m_{B_{s}} }\,
     A_{2}^{ B_{c}^{\ast} {\to} B_{s} }
     \label{bs-rho-d}, \\
    {\cal M}_{p}^{ B_{s}{\rho} } &=&
    -\;\,\frac{ G_{F} }{ \sqrt{2} }\, V_{cs}^{\ast}\,V_{ud}\,
     f_{\rho}\, m_{\rho}\, a_{1}\,
     \frac{ 2\,m_{B_{c}^{\ast}}\,m_{\rho} }
          { m_{B_{c}^{\ast}}+m_{B_{s}} }\,
     V^{ B_{c}^{\ast} {\to} B_{s} }
     \label{bs-rho-p},
     \end{eqnarray}
     \begin{equation}
    {\cal M}_{i}^{ B_{s}K^{\ast} } \, =\,
    {\cal M}_{i}^{ B_{s}{\rho} }
     \big( V_{ud}{\to}V_{us},
     \, f_{\rho}{\to}f_{K^{\ast}},
     \, m_{\rho}{\to}m_{K^{\ast}} \big),
     \quad \text{for } i=s,p,d
     \label{bs-kv}.
     \end{equation}

   For $B_{c}^{{\ast}+}$ ${\to}$ $B_{d}^{0}{\rho}^{+}$,
   $B_{d}^{0}K^{{\ast}+}$ decays, one has
     \begin{eqnarray}
    {\cal M}_{s}^{ B_{d}{\rho} } &=&
     -i\, \frac{ G_{F} }{ \sqrt{2} }\, V_{cd}^{\ast}\,V_{ud}\,
     f_{\rho}\, m_{\rho}\, a_{1}\,
     \big( m_{B_{c}^{\ast}}+m_{B_{d}} \big)\,
     A_{1}^{ B_{c}^{\ast} {\to} B_{d} }
     \label{bd-rho-s}, \\
    {\cal M}_{d}^{ B_{d}{\rho} } &=&
     -i\, \frac{ G_{F} }{ \sqrt{2} }\, V_{cd}^{\ast}\,V_{ud}\,
     f_{\rho}\, m_{\rho}\, a_{1}\,
     \frac{ 2\,m_{B_{c}^{\ast}}\,m_{\rho} }
          { m_{B_{c}^{\ast}}+m_{B_{d}} }\,
     A_{2}^{ B_{c}^{\ast} {\to} B_{d} }
     \label{bd-rho-d}, \\
    {\cal M}_{p}^{ B_{d}{\rho} } &=&
    -\;\,\frac{ G_{F} }{ \sqrt{2} }\, V_{cd}^{\ast}\,V_{ud}\,
     f_{\rho}\, m_{\rho}\, a_{1}\,
     \frac{ 2\,m_{B_{c}^{\ast}}\,m_{\rho} }
          { m_{B_{c}^{\ast}}+m_{B_{d}} }\,
     V^{ B_{c}^{\ast} {\to} B_{d} }
     \label{bd-rho-p},
     \end{eqnarray}
     \begin{equation}
    {\cal M}_{i}^{ B_{d}K^{\ast} } \, =\,
    {\cal M}_{i}^{ B_{d}{\rho} }
     \big( V_{ud}{\to}V_{us},
     \, f_{\rho}{\to}f_{K^{\ast}},
     \, m_{\rho}{\to}m_{K^{\ast}} \big),
     \quad \text{for } i=s,p,d
     \label{bd-kv}.
     \end{equation}

   For $B_{c}^{{\ast}+}$ ${\to}$ ${\eta}_{c}{\rho}^{+}$,
   ${\eta}_{c}K^{{\ast}+}$ decays, one has
     \begin{eqnarray}
    {\cal M}_{s}^{ {\eta}_{c}{\rho} } &=&
     -i\, \frac{ G_{F} }{ \sqrt{2} }\, V_{cb}^{\ast}\,V_{ud}\,
     f_{\rho}\, m_{\rho}\, a_{1}\,
     \big( m_{B_{c}^{\ast}}+m_{{\eta}_{c}} \big)\,
     A_{1}^{ B_{c}^{\ast} {\to} {\eta}_{c} }
     \label{etac-rho-s}, \\
    {\cal M}_{d}^{ {\eta}_{c}{\rho} } &=&
     -i\, \frac{ G_{F} }{ \sqrt{2} }\, V_{cb}^{\ast}\,V_{ud}\,
     f_{\rho}\, m_{\rho}\, a_{1}\,
     \frac{ 2\,m_{B_{c}^{\ast}}\,m_{\rho} }
          { m_{B_{c}^{\ast}}+m_{{\eta}_{c}} }\,
     A_{2}^{ B_{c}^{\ast} {\to} {\eta}_{c} }
     \label{etac-rho-d}, \\
    {\cal M}_{p}^{ {\eta}_{c}{\rho} } &=&
    -\;\,\frac{ G_{F} }{ \sqrt{2} }\, V_{cb}^{\ast}\,V_{ud}\,
     f_{\rho}\, m_{\rho}\, a_{1}\,
     \frac{ 2\,m_{B_{c}^{\ast}}\,m_{\rho} }
          { m_{B_{c}^{\ast}}+m_{{\eta}_{c}} }\,
     V^{ B_{c}^{\ast} {\to} {\eta}_{c} }
     \label{etac-rho-p},
     \end{eqnarray}
     \begin{equation}
    {\cal M}_{i}^{ {\eta}_{c}K^{\ast} } \, =\,
    {\cal M}_{i}^{ {\eta}_{c}{\rho} }
     \big( V_{ud}{\to}V_{us},
     \, f_{\rho}{\to}f_{K^{\ast}},
     \, m_{\rho}{\to}m_{K^{\ast}} \big),
     \quad \text{for } i=s,p,d
     \label{etac-kv}.
     \end{equation}

  \end{appendix}


  \end{document}